# Novel Fiber Design for Wideband Conversion and Amplification in Multimode Fibers

M.Guasoni[(1)], F.Parmigiani[(1)], P.Horak[(1)], D.J.Richardson[(1)]

[(1)] Optoelectronics Research Centre, University of Southampton, Southampton, SO17 1BJ, United Kingdom, m.guasoni@soton.ac.uk

**Abstract** *We propose an operating principle to achieve broadband and highly tunable mode conversion and amplification exploiting inter-modal four wave mixing in a multimode fiber. A bandwidth of 30 nanometers is demonstrated by properly designing a simple step-index silica fiber.*

## Introduction

Space-Division-Multiplexing (SDM) in multimode (MM) fibers has rapidly emerged over the past few years as one of the most promising solutions to the anticipated capacity crunch of single-mode fiber systems[1]. In this framework, the development of broadband multimode inline amplifiers and mode/wavelength converters for long-haul SDM transmissions is of utmost importance. Mode and wavelength conversion as well as amplification can be achieved by exploiting intermodal (IM-) four-wave-mixing (FWM) in a multimode optical fiber. IM-FWM paves the way to all-fiber multimode amplifiers and wavelength converters whose operating frequencies can be tuned over a far broader spectrum than in conventional MM-Raman amplifiers and MM-Erbium-doped-fiber-amplifiers, and be tuned to operate far away from the spectral bandwidth of the main noise sources (spontaneous Raman scattering and amplified spontaneous emission) that typically impair FWM in single-mode fibers[2].

Among the two main IM-FWM processes, namely Bragg-Scattering (BS) and Phase-Conjugation (PC), only PC leads to exponential amplification of the corresponding spatial modes, together with the simultaneous mode/wavelength conversion. Some recent works[2-3] proposed the capability of achieving broadband operation for the BS process when properly designing the inverse group velocity (IGV) curves of the corresponding modes. However, for these fiber designs the PC process exhibits an operating bandwidth of less than one nanometer, preventing its practical usage for mode amplification.

In this work we propose a new fiber design to allow broadband operation of the PC process and demonstrate up to several tens of nanometer bandwidth even in a simple step-index silica fiber.

These results can pave the way for the development of a new class of wavelength tunable, wideband multimode converters and amplifiers.

## Basic principle for wideband PC operation

The configuration of the BS and PC processes and corresponding waves is presented in Fig.1 for the simple case of exciting only two modes. Note that the configuration can be easily extended to multiple modes. Two input pumps $P_0$ and $P_1$ are coupled to two distinct spatial modes of the fiber that we indicate with $m_0$ and $m_1$, respectively. An input seed signal $S_0$ is coupled to mode $m_0$; due to IM-FWM an idler wave is generated in $m_1$ for both the BS ($I_{1,BS}$) and the PC ($I_{1,PC}$) process. The pumps are centered at the frequencies $\omega_{p0}$ and $\omega_{p1}$, respectively, and the signal at the frequency $\omega_{s0} = \omega_{p0} + \Delta\omega$, where $\Delta\omega$ is the pump-signal detuning. According to energy conservation, the idler replica are generated at frequency $\omega_{i1} = \omega_{p1} + \Delta\omega$ for the BS process and at frequency $\omega_{i1} = \omega_{p1} - \Delta\omega$ for the PC process, respectively.

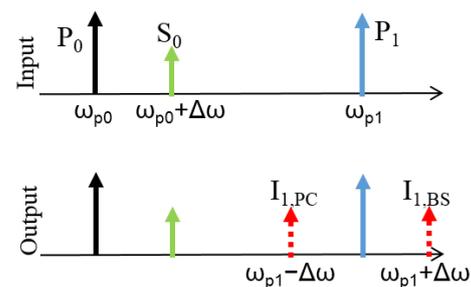

**Fig. 1:** Representation of the Bragg Scattering and Phase-Conjugation processes. The input waves $P_0$, $P_1$ and $S_0$ generate the idler component $I_{1,BS}$ (Bragg scattering) and $I_{1,PC}$ (Phase-Conjugation).

The idler generation is maximized when the phase mismatch $\Delta\beta$ is almost null[2,3]. For BS processes, this condition is achieved over a wide bandwidth whenever the IVG of one mode is a shifted copy of the other by $\omega_{p1} - \omega_{p0}$ (see Fig.2a)[2,3]. The phase mismatch for the PC process is[2,3] $\Delta\beta = \beta^{(0)}(\omega_{p0}) + \beta^{(1)}(\omega_{p1}) - \beta^{(1)}(\omega_{i1}) - \beta^{(0)}(\omega_{s0})$, where $\beta^{(0)}(\omega)$ and $\beta^{(1)}(\omega)$ indicate the propagation constant of modes $m_0$ and $m_1$ at the frequency $\omega$, respectively. Consequently, the condition Δβ = 0 is achieved whenever $\beta^{(0)}(\omega_{p0}+\Delta\omega) - \beta^{(0)}(\omega_{p0}) = \beta^{(1)}(\omega_{p1}) - \beta^{(1)}(\omega_{p1}-\Delta\omega)$. The difference $\beta^{(i)}(\omega_b) - \beta^{(i)}(\omega_a)$ equals the definite integral in the interval [$\omega_a$, $\omega_b$] of the derivative function $\beta_1^{(i)}=\partial\beta^{(i)}/\partial\omega$, which is the IGV of mode $m_i$. The condition Δβ = 0 can thus be rewritten as follows:

$$\int_{\omega_{p0}}^{\omega_{p0}+\Delta\omega}\beta_1^{(0)}\partial\omega = \int_{\omega_{p1}-\Delta\omega}^{\omega_{p1}}\beta_1^{(1)}\partial\omega \quad (1)$$

Note that the condition Eq.(1) is fulfilled for any detuning Δω if:

$$\beta_1^{(0)}(\omega_c+\omega) = \beta_1^{(1)}(\omega_c-\omega) \quad (2)$$

that is to say, when the IGVs $\beta_1^{(0)}$ and $\beta_1^{(1)}$ exhibit a mirror symmetry with respect to $\omega_c=(\omega_{p0}+\omega_{p1})/2$, as displayed in Fig.2b. Equation (2) represents the optimal condition to achieve broadband PC operation. It follows that the odd dispersion terms of the two modes computed at $\omega_c$ are equal, that is $\beta_{2n+1}^{(0)}(\omega_c)= \beta_{2n+1}^{(1)}(\omega_c)$, whereas the even dispersion terms are opposite, that is $\beta_{2n}^{(0)}(\omega_c)= -\beta_{2n}^{(1)}(\omega_c)$, where $\beta_p=\partial^p\beta/\partial\omega^p$.

Therefore, the conditions required on the IGVs of the two modes are completely different for the two instances in which broadband PC operation or broadband BS operation are targeted. Consequently, tailored fiber engineering is necessary to achieve a wide bandwidth in the two cases.

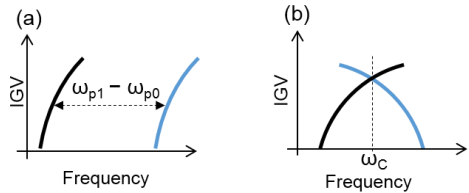

**Fig. 2:** (a) Optimal condition for broadband BS operation: the IGVs of the two modes $m_0$ (black) and $m_1$ (blue) are shifted copies of each other. The shift corresponds to the pump-to-pump detuning $\omega_{p1}-\omega_{p0}$. (b) Optimal condition for broadband PC operation: the two modal IGVs exhibit a mirror symmetry centered at $\omega_c=(\omega_{p0}+\omega_{p1})/2$.

**Fiber design**
Past works have focused on BS processes and pointed out some fiber configurations where a BS bandwidth of some nanometers[2,3] is achievable at telecom wavelengths. On the other hand, for the same configurations, a PC bandwidth below one nanometer and with very low conversion efficiency is predicted. The fibers considered in those works are graded-index fibers whose modes $LP_{01}$ and $LP_{11}$ possess IGVs that are almost parallel over some nanometers bandwidth in the C-band, so that the optimal condition for the BS process illustrated in Fig.2a is fulfilled.

On the contrary, in order to achieve a broadband PC operation the IGVs should exhibit a mirror symmetry, as displayed in Fig.2b. Here we focus on step-index circular core silica fibers, which represent a practical platform in experiments. Moreover, they are characterized by just two degrees of freedom, namely the core radius R and the core-cladding index difference Δn, which allows for relatively fast and straightforward computation of the most suitable configurations leading to broadband PC operation. On the other hand, having two degrees of freedom means that we can only impose the two conditions $\beta_1^{(0)}(\omega_c)= \beta_1^{(1)}(\omega_c)$ and $\beta_2^{(0)}(\omega_c)= -\beta_2^{(1)}(\omega_c)$, without control over the higher-order dispersion terms.

Interestingly, we find that for typical values 6 µm ≤ R ≤ 12 µm and 0.002 ≤ Δn ≤ 0.008 the two aforementioned conditions can be fulfilled when the modes $LP_{01}$ and $LP_{02}$ are employed. As an example, Fig.3 illustrates the IGVs of the two modes in a fiber with R=8.01 µm and Δn = 0.0063. At the telecom wavelength $\lambda_c=2\pi c/\omega_c=1531.1$ nm the two IGVs intersect ($\beta_1^{(0)}(\lambda_c)=\beta_1^{(1)}(\lambda_c)=4898.4$ ps/m) and have opposite slope ($\beta_2^{(0)}(\lambda_c)= -\beta_2^{(1)}(\lambda_c)= -27.67$ ps$^2$/km).

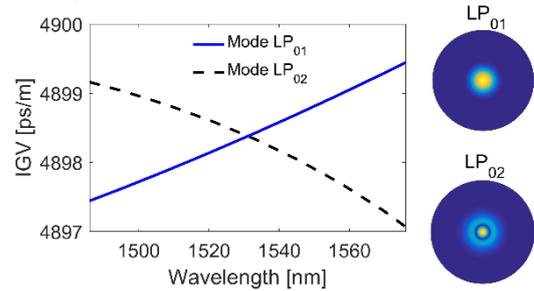

**Fig. 3:** IGV of modes LP01 and LP02 in a step-index silica fiber with R=8.01 µm and Δn = 0.0063. The two IGVs intersect at $\lambda_c$ = 1531.1 nm, where they have opposite slope. A mirror symmetry between the two IGVs is therefore achieved around this wavelength. On the left, the transverse mode profiles are displayed.

Therefore, with good approximation the IGVs of the two modes are characterized by mirror symmetry in a frequency band close to $\lambda_c$. Far from $\lambda_c$ the symmetry is not preserved because the higher-order dispersion coefficients of the two modes do not satisfy the conditions $\beta_{2n+1}^{(0)}(\lambda_c)= \beta_{2n+1}^{(1)}(\lambda_c)$ and $\beta_{2n}^{(0)}(\lambda_c)=-\beta_{2n}^{(1)}(\lambda_c)$. For example, the third-order coefficients for the two modes are $\beta_3^{(0)}(\lambda_c)=0.14$ ps$^3$/km and $\beta_3^{(1)}(\lambda_c)= -0.41$ ps$^3$/km, respectively. The intramodal effective areas of modes $LP_{01}$ and $LP_{02}$ are 151 µm$^2$ and 224 µm$^2$, respectively, whereas the intermodal effective area accounting for the overlap among modes $LP_{01}$ and $LP_{02}$ is 251 µm$^2$. These areas are almost constant across a band of several tens of nanometers around $\lambda_c$. We simulate the PC process in the aforementioned fiber by solving the MM-Nonlinear Schrödinger Equation (NLSE) (see e.g. Ref[2]) for the system parameters discussed above. Pumps and signal are linearly copolarized at the input. The input power of the seeding signal is -17 dBm.

Figure 4a displays the Idler Conversion Efficiency (ICE, namely the power ratio between the output idler and the input signal) as function of the pump ($P_0$)-to-signal ($S_0$) detuning $\Delta\lambda_{PS}$ and the pump-to-pump detuning $\Delta\lambda_{PP}$ when the fiber length is L=1 km and the power of both input pumps is 30 dBm. We find that for a pump-to-pump detuning $\Delta\lambda_{PP}$=0.080 nm the largest operational bandwidth ( at – 3 dB) is achieved, which is about 9 nm with a peak conversion efficiency of 10.8 dB (see Fig. 4b). On the other hand, we notice that the ICE changes remarkably when a variation of $\Delta\lambda_{PP}$ as small as 0.078 nm (10 GHz) is applied. This suggests that the system dynamics may be strongly sensitive to small inaccuracies in the pump-to-pump detuning and to weak random variations of the fiber parameters that typically occur on a length scale of a few meters[4].

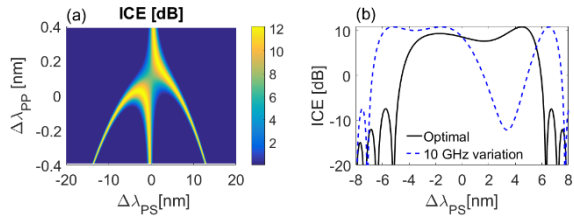

**Fig. 4:** (a) ICE versus $\Delta\lambda_{PP}$ and $\Delta\lambda_{PS}$ when L=1 km and the input pump powers are 30 dBm. (b) ICE versus $\Delta\lambda_{PS}$ at the optimal pump-to-pump detuning $\Delta\lambda_{PP}$=0.080 nm (black solid line) and for $\Delta\lambda_{PP}$=0.002 nm (blue dotted line), corresponding to a small variation of 10 GHz with respect to the optimal detuning.

In order to make the system robust against these inaccuracies and random variations, we increase the input pump power by a factor of 50 (47 dBm) and we reduce the fiber length by the same factor (L=20 m), so that the total degree of nonlinearity is preserved.

This induces a double benefit: first, the fiber length is strongly reduced, so that random longitudinal variations of the fiber play a minor role; second, the large input power induces an effective nonlinear contribution to the phase-mismatch, which results in a broadening of the ICE along both the $\Delta\lambda_{PP}$ axis and the $\Delta\lambda_{PS}$ axis, as illustrated in Fig.5a. As shown in Fig.5b, in this case the largest PC bandwidth at -3dB is 30 nm, achieved for $\Delta\lambda_{PP}$ = -0.24 nm, whereas the ICE peak is still 10.8 dB because the total nonlinear length is preserved. Furthermore, in this configuration the system is robust against a variation of the detuning $\Delta\lambda_{PP}$ of several tens of GHz (see Fig.5b). Note that the PC bandwidth and the system robustness may be increased by further increasing the input power and by reducing the fiber length up to a few meters.

It is worth mentioning that in our system the central wavelength $\lambda_c$ can be tuned in the whole band from 1400 nm to 1600 nm by properly adjusting R and $\Delta n$ within the range of values previously mentioned (µm ≤ R ≤ 12 µm and 0.002 ≤ $\Delta n$ ≤ 0.008).

We conclude by noting that the usage of a highly nonlinear fiber may replace the necessity of high-power sources.

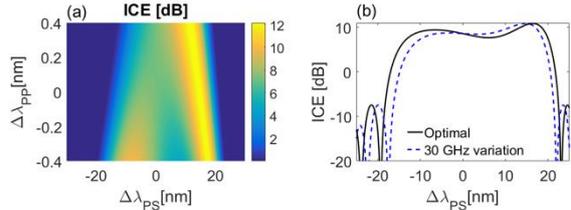

**Fig. 5:** Same as in Fig.4 but when L=20 m and the input pump powers are 47 dBm. The optimal detuning $\Delta\lambda_{PP}$ = -0.24 nm. The dotted-blue curve corresponds to a 30 GHz variation with respect to the optimal detuning.

**Conclusions**

We have proposed and demonstrated numerically wideband phase conjugation operation in a step-index multimode fiber. The fiber was designed such that the inverse group velocities of the two fiber modes exhibit mirror symmetry. An operating bandwidth of several tens of nanometers is obtained when modes $LP_{01}$ and $LP_{02}$ are employed. We have suggested that shortening the fiber length to a few meters (and thus increasing the pump power or the fiber nonlinearity to maintain the same degree of overall nonlinearity) makes the PC process robust against inaccuracies of the system parameters and longitudinal random variations along the fiber length.

**Acknowledgements**
M. Guasoni is supported through an Individual Marie Sklodowska-Curie Fellowship (H2020 MSCA IF 2015, project AMUSIC - Grant Agreement 702702).